\documentclass[notitlepage,aps,prl,twocolumn,superscriptaddress,amsmath,showpacs,tightenlines,longbibliography,10pt,footnote]{revtex4-1}

\usepackage{amssymb}
\usepackage{amsmath}
\usepackage{dcolumn}
\usepackage{graphicx}
\usepackage{mathrsfs}
\usepackage{subfigure}
\usepackage{booktabs}
\usepackage{color}
\usepackage{docmute}
\usepackage{hyphenat}
\usepackage{CJKutf8}
\usepackage{multirow}
\usepackage{array}
\usepackage{threeparttable}
\usepackage[mathscr]{euscript}

\usepackage{graphicx}
\usepackage{subfigure}
\usepackage{tikz}
\usetikzlibrary{positioning}
\usepackage{diagbox}

\newcommand{\bea}{\begin{eqnarray}}
\newcommand{\eea}{\end{eqnarray}}
\newcommand{\beq}{\begin{equation}}
\newcommand{\eeq}{\end{equation}}
\newcommand{\nn}{\nonumber}
\def\/{\over}

\usepackage{url}
\usepackage[colorlinks]{hyperref}
\hypersetup{%
	plainpages=true,
	breaklinks=true,
	hypertexnames=false,
	pageanchor=true,
	bookmarksnumbered=true,
	colorlinks=true,
	linkcolor={blue},
	citecolor={red},
	urlcolor={blue},
	anchorcolor={black}
}

\allowdisplaybreaks[4]

\begin{document}
\title{Significant circular Unruh effect at small acceleration}
\author{Yuebing Zhou}
\affiliation{Department of Physics, Huaihua University, Huaihua, Hunan 418008, China}
\affiliation{Department of Physics, Synergetic Innovation Center for Quantum Effects and Applications,\\ and Institute of Interdisciplinary Studies, \\Hunan Normal University, Changsha, Hunan 410081, China}
\author{Jiawei Hu}
\email[Corresponding author: ]{jwhu@hunnu.edu.cn}
\affiliation{Department of Physics, Synergetic Innovation Center for Quantum Effects and Applications,\\ and Institute of Interdisciplinary Studies, \\Hunan Normal University, Changsha, Hunan 410081, China}
\author{Hongwei Yu}
\email[Corresponding author: ]{hwyu@hunnu.edu.cn}
\affiliation{Department of Physics, Synergetic Innovation Center for Quantum Effects and Applications,\\ and Institute of Interdisciplinary Studies, \\Hunan Normal University, Changsha, Hunan 410081, China}

\begin{abstract}
We study the transition rates of an atom rotating in a circular orbit, which is coupled with fluctuating electromagnetic fields in vacuum. We find that when the rotational angular velocity exceeds the transition frequency of the atom, the excitation rate can reach the same order of magnitude as the emission rate,  even with an extremely low centripetal acceleration resulting  from a very small orbital radius. For experimentally accessible centripetal accelerations, the excitation rate of centripetally accelerated atoms can be $10^{272,878}$ times greater than that of linearly accelerated atoms with the same  magnitude of acceleration. Our result suggests that the circular version of the Unruh effect can be significant even at very small centripetal accelerations, contrary to the common belief that a large Unruh effect requires large acceleration. This finding sheds new light on the experimental detection of the circular Unruh effect.

\end{abstract}
\maketitle

\emph{Introduction}.---It is well known that, in vacuum, an atom in an excited state can spontaneously transition to its ground state and emit a photon, whereas an atom in the ground state can never transition to an excited state. Remarkably, however, spontaneous excitation can occur for uniformly accelerated ground-state atoms, and the spontaneous excitation rate $\Gamma_{-}$ is related to the spontaneous emission rate $\Gamma_{+}$ as
\bea\label{ratio}
{\Gamma_-}=e^{-{2\pi\omega_0}/{a}}{\Gamma_+}=e^{-{\omega_0}/{T_{U}}}{\Gamma_+},
\eea
where $\omega_0$ and $a$ are the transition frequency and proper acceleration of the atom respectively. This means that a uniformly accelerated atom would excite as if it were immersed in a  bath of thermal radiation at a temperature $T_{U}=a/2\pi$. 
This striking phenomenon is known as the Unruh effect \cite{Fulling1973,Davies1975,W. G. Unruh} (see Refs. \cite{Birrell1982,Wald1994,Crispino2008}, for reviews). The Unruh effect is not only  intriguing on its own but also closely related to other quantum effects in curved spacetimes, such as the Hawking radiation \cite{Hawking74}, from the point of view of the equivalence principle regarding acceleration and gravity. 
Equation~\eqref{ratio} reveals that the spontaneous excitation rate of a uniformly accelerated atom will be much smaller than the spontaneous emission rate when the acceleration is small compared with the transition frequency of the atom. This indicates that a very large acceleration  compared with the atomic transition frequency  is required in experimental detection of the effect, which is extremely challenging. 

One may wonder what happens if an atom experiences  centripetal acceleration, which is another common type of acceleration besides uniform acceleration. In this context, it has been found that a centripetally accelerated observer, similar to a uniformly accelerated one, would also perceive radiation in a vacuum, although the spectrum perceived is nonthermal \cite{Bell83,Hacyan1986,Bell87,Kim1987,Unruh98,Rosu2005}. Correspondingly, atoms undergoing centripetal acceleration can also get excited spontaneously \cite{Letaw1981,Korsbakken2004,Good2020,Takagi1984}, 
although the relation between the excitation rate and the emission rate is now not as simple as Eq.~\eqref{ratio}. This phenomenon is dubbed the circular Unruh effect. Intuitively, one may expect that a large centripetal acceleration is also necessary to obtain a relatively large spontaneous excitation rate. 
In fact,  extensive studies on the circular Unruh effect over recent decades align with this expectation~\cite{Letaw1980,Rogers88,Lochan20,Biermann2020,Takagi1986,YaoJin2014-1,YaoJin2014-2,Bunney23,Bunney23-2,Gim2018,Levin1993,Davies1996,Lorenci2000}.  Many of these studies were conducted in the ultrarelativistic limit, i.e., in the limit where the linear velocity approaches the speed of light while the centripetal acceleration remains finite~\cite{Takagi1984, Kim1987, Takagi1986, YaoJin2014-1, YaoJin2014-2, Biermann2020, Bunney23}. In this limit, the relation between the upward and downward transition rates is similar to that in the linear acceleration case described by Eq.~\eqref{ratio}, and an effective temperature can thus be defined  using the ratio between these two rates, which is approximately proportional to the centripetal acceleration. 
There are also investigations in which the ultrarelativistic limit is not taken \cite{Letaw1980,Rogers88,Levin1993,Davies1996,Lorenci2000,Gim2018,Lochan20,Bunney23-2}. 
However, to the best of our knowledge, in all known works, the spontaneous excitation rate of centripetally accelerated atoms approaches zero in the limit of vanishingly small centripetal acceleration. This is consistent with what we have learned from the linear Unruh effect and seems to imply that an extremely large centripetal acceleration is indeed necessary to detect the circular Unruh effect with spontaneous excitation.

Contrary to the common wisdom that a large Unruh effect demands a large acceleration, we discover in this Letter that an atom rotating in a circular orbit can exhibit a large spontaneous excitation rate even at a vanishingly small centripetal acceleration. 
Specifically, we investigate the transition rates of a centripetally accelerated atom coupled with fluctuating electromagnetic fields in vacuum. We show in detail that when the angular velocity exceeds the transition frequency of the atom, the spontaneous excitation rate, even in the limit of a vanishingly small orbital radius and consequently a vanishingly small centripetal acceleration, is not only nonvanishing but also, unexpectedly, of the same order of magnitude as the emission rate.
Moreover, the effective temperature defined by the ratio of the excitation and emission rates is approximately proportional to the rotational angular velocity when the angular velocity is much larger than the transition frequency of the atom. This is remarkably different from the linear acceleration case, where the temperature is proportional to the acceleration. This suggests that we can observe a significant circular Unruh effect at an almost vanishing centripetal acceleration. 
Units with $\hbar=c=\epsilon_0=k_B=1$ are used in this Letter, where $\hbar$ is the reduced Planck constant, $c$ the speed of light, $\epsilon_0$ the vacuum permittivity, and $k_B$ the Boltzmann constant.

\emph{The transition rates for a rotating atom}.---We consider a circularly moving two-level atom in interaction with the fluctuating electromagnetic fields in vacuum. The atom is assumed to be rotating at an angular velocity $\Omega$ in a circular orbit with a radius $R$, and so its trajectory can be described in the cylindrical coordinates as
\begin{eqnarray}\label{trajectories}
&&
r(t)=R,\;\;\;\;\;\;\theta(t)=\Omega t,\;\;\;\;\;\;z(t)=0,
\end{eqnarray}
where $t$ is the coordinate time. 
The atom is assumed to be polarizable but does not possess a permanent dipole moment. 
However, it can be instantaneously polarized by fluctuating electromagnetic fields in vacuum. 
In the proper frame of the atom, the Hamiltonian describing the atom-field interaction can be written in a manifestly invariant form as \cite{Takagi1986} $H_I^{(\text{0})}=-D^{\mu}F_{\mu\nu}u^{\nu}$, where $F_{\mu\nu}$ is the electromagnetic tensor, $D^{\mu}$ is the four-electric dipole moment, which can be expressed as $(0,D_r,D_{\theta},D_z)$ in the proper frame, and $u^{\nu}$ is the four velocity. 
In the following, we work in the laboratory frame, in which $D^{\mu}=(\gamma R\Omega D_{\theta},D_r,\gamma D_{\theta}/R,D_z)$, and $u^{\nu}=(\gamma,0,\gamma\Omega,0)$. 
Thus, the interaction Hamiltonian in the laboratory frame can be written as
\bea\label{hi2}
H_I=-\sum_{i=r,\theta,z}D_{i}\mathcal{E}_{i}(t,{\bf x}),
\eea
where
\bea\label{hi3}
&&\mathcal{E}_{r}(t,{\bf x})=E_{r}(t,{\bf x})+R \Omega B_{z}(t,{\bf x}), \nn\\
&&\mathcal{E}_{\theta}(t,{\bf x})=\gamma^{-1}E_{\theta}(t,{\bf x}),\nn\\
&&\mathcal{E}_{z}(t,{\bf x})=E_{z}(t,{\bf x})-R \Omega B_{r}(t,{\bf x}).
\eea
Here
$E_{i}(t,{\bf x})$ and $B_{i}(t,{\bf x})$ are the electric field and magnetic induction strength, respectively, and $(t,{\bf x})$ is the abbreviation for $(t,r,\theta,z)$.

In the framework of open quantum systems, the emission rate $\Gamma_+$ and the excitation rate $\Gamma_-$ can be derived as, 
\bea\label{eer}
\Gamma_{\pm}(\omega)=\sum_{i,j=r,\theta,z}d_{i}d_{j}^{\ast}\int^{\infty}_{-\infty}G_{ij}(\Delta t)\; e^{\pm i\omega\Delta t} d\Delta t,
\eea
where $d_{i}=\langle1|D_{i}|0\rangle$ is the dipole transition matrix element, $\omega$ is the energy level spacing of the atom in the laboratory frame, which is related to that in the proper frame of the atom $\omega_0$ via $\omega=\omega_0/\gamma>0$, and
\bea\label{field correlation function}
&G_{ij}(\Delta t)=\big\langle\mathcal{E}_{i}\big(t,{\bf x}(t)\big)\mathcal{E}_{j}\big(t',{\bf x}(t')\big)\big\rangle
\eea
are the electromagnetic field correlation functions, with $\langle\cdot\cdot\cdot\rangle$ denoting the expectation value with respect to the vacuum state of the electromagnetic field. Note that the field correlation functions here are invariant under temporal translations, i.e., they are functions of $\Delta t=t-t'$. 
See Sec.~\textcolor{blue}{I} of the Supplemental Material for the derivation of Eq. \eqref{eer}.
For simplicity, we assume that the transition matrix elements $d_{i}$ are real, and then the cross terms in Eq.~\eqref{eer} will be vanishing. 
By substituting  the trajectory of the atom Eq. \eqref{trajectories} into the general form of the field correlation functions shown in Eq. \eqref{field correlation function},  
and with the help of the quantized vector potential of the electromagnetic field in the cylindrical coordinates \cite{Aliev1989,HCai2015}, 
one obtains the explicit forms of the field correlation functions in the laboratory frame as
\begin{widetext}{\center}
\bea
G_{rr}(\Delta t)&=&\frac{1}{4 \pi ^2}\sum _{m=-\infty }^{+\infty }\,\sum_{\zeta,\alpha,\beta=-1}^{1}\int _0^{+\infty }d k_{\bot}\int _{-\infty }^{+\infty }d k_z\frac{k_{\bot}}{4 \omega_k}J_{\left| m+\zeta\right|+\alpha}(k_{\bot} R) J_{\left| m+\zeta\right|+\beta}\left(k_{\bot} R\right)\,e^{i(m\Omega-\omega_k)\Delta t}\hspace{0.5cm}\nn\\
&&\;\;\;\;\;\;\;\;\;\;\times
\left\{\frac{| \zeta |  \left(R^2 \Omega ^2+2\right)-2}{4}k_{\bot}^2\alpha \beta+ (-1)^{\frac{|\alpha|+|\beta| +\alpha +\beta}{2}}\frac{ | \zeta | (1-| \alpha \beta| ) (R\Omega k_{\bot})^{|\alpha|+|\beta|}}{2[\Omega+\zeta (m\Omega-\omega_k)]^{|\alpha|+|\beta|-2}}\right\},\label{Wightman-correlation-function-rr-m}\\
G_{\theta\theta}(\Delta t)
&=&\frac{1}{4 \pi ^2}\sum _{m=-\infty }^{+\infty }\,\sum_{\zeta=-1}^{1}\int _0^{+\infty }d k_{\bot}\int _{-\infty }^{+\infty }d k_z\frac{k_{\bot}}{4 \omega_k}\,J^2_{\left| m+\zeta\right|}(k_{\bot} R)\,e^{i(m\Omega-\omega_k)\Delta t}
\gamma^{-2}\left[\frac{2m^2}{R^2}(|\zeta|-1)+\omega_{k}^{2}|\zeta|\right] ,\label{Wightman-correlation-function-theta-m}\hspace{1cm}
\\
G_{zz}(\Delta t)&=&\frac{1}{4 \pi ^2}\sum _{m=-\infty }^{+\infty }\,\sum_{\zeta=-1}^{1}\int _0^{+\infty }d k_{\bot}\int _{-\infty }^{+\infty }d k_z\frac{k_{\bot}}{4 \omega_k}J^2_{\left| m+\zeta\right| }(k_{\bot} R)\,e^{i(m\Omega-\omega_k)\Delta t}
\hspace{0.5cm}\nn\\&&\;\;\;\;\;\;\;\;\;\;\times
\left\{2
(m\Omega-\omega_k)^2(1-|\zeta|)
+k_z^2\left[(3
-\gamma^{-2})|\zeta|-2
\right]\right\},\label{Wightman-correlation-function-zz-m}\hspace{1cm}
\eea
where $J_\nu(z)$ is the Bessel function of the first kind. See Sec.~\textcolor{blue}{II} of the Supplemental Material for the derivation of Eqs. \eqref{Wightman-correlation-function-rr-m}-\eqref{Wightman-correlation-function-zz-m}. 
Plugging the explicit forms of the field correlation functions, Eqs.~\eqref{Wightman-correlation-function-rr-m}-\eqref{Wightman-correlation-function-zz-m}, into Eq. \eqref{eer}, the transition rates  $\Gamma_{\pm}$ can be found to be 
\bea\label{Gammapm}
\Gamma_{\pm}(\omega)&=&\sum _{m=-\infty}^{+\infty }\frac{(m\Omega\pm\omega)^3}{3\pi}\Theta (m \Omega \pm\omega )P_m^{\pm}(\Omega/\omega,R\omega),\hspace{0.5cm}
\eea
where $\Theta(x)$ the Heaviside theta function which equals  $0$ for $x<0$ and $1$ for $x>0$, and
\bea
P_m^{\pm}(\Omega/\omega,R\omega)=\sum _{\zeta,\alpha,\beta}\sum_{k=0}^{+\infty } \frac{3\left(d_{r}^2 \mathcal{M}^{\pm}_{m}+d_{\theta}^2 \mathcal{N}^{\pm}_{m}+d_{z}^2 \mathcal{Z}^{\pm}_{m}\right)(-1)^{k+h} \left[\left(k+p+q\right)!\right]^2(2 k+2 p)![R\omega (m \Omega/\omega \pm1 )]^{2(k+p+q)}}{4\left(\big|| m+\zeta |+\alpha\big| +k\right)! \left(\big|| m+\zeta |+\beta\big| +k\right)! (2 k+2 p+2q +1)!(k+2 p)! k! }\,.\hspace{0.5cm}
\eea
Here, $\zeta,\alpha,\beta\in\{-1,0,1\}$, $p=\frac{\big|| m+\zeta |+\alpha\big|+\big|| m+\zeta |+\beta\big|}{2}$, $q=\frac{|\alpha|+|\beta|}{2}$, $g=\frac{\alpha+\beta}{2}$, $h=p-g-| m+\zeta |$, and 
\bea
\mathcal{M}^{\pm}_{m}&=&
\frac{(-1)^{q+g} | \zeta | (1-| \alpha \beta| ) (\Omega/\omega)^{2q}}{(\Omega/\omega \mp\zeta )^{2q-2}(m\Omega/\omega \pm1 )^{2}}+\frac{ | \zeta |  \left[(R\omega)^2 (\Omega/\omega)^2+2\right]-2}{(R\omega)^2(m\Omega/\omega \pm1 )^{2}}\alpha \beta
,\\
\mathcal{N}^{\pm}_{m}&=&\left[| \zeta |-\frac{2 m^2 (1-| \zeta | )}{(R\omega)^2 (m \Omega/\omega \pm1 )^{2}}\right]\frac{(1-| \alpha| )(1-|\beta| )}{\gamma^2},\hspace{0.5cm}\\
\mathcal{Z}^{\pm}_{m}&=&\left[\frac{2
(1-\left| \zeta \right| )}{(m \Omega/\omega\pm1)^2}+\frac{\left(3
-\gamma^{-2}\right) \left| \zeta \right| -2
}{2 \left| m+\zeta \right| +2 k+3}\right](1-| \alpha| )(1-|\beta| ).
\eea
See Sec.~\textcolor{blue}{III} of the Supplemental Material for the derivation of Eq. \eqref{Gammapm}. In the following, we will separately calculate the contributions to the transition rates of the polarization of the atom along the axis of rotation and that perpendicular to it.
\end{widetext}

\emph{(I)\;Atomic polarization along the axis of rotation}:
First, we consider the transition rate contributed by the atomic polarization along the direction of the rotation axis (i.e., the $z$-axis). When the orbital radius $R$ is small such that $v=R\Omega\ll1$ and $R\omega_0\ll1$, the emission and excitation rates of the atom Eq. \eqref{Gammapm} can be approximated as
\bea\label{approximate-z}
\Gamma_{+}&=&\frac{d_{z}^2\omega _0^3}{3 \pi }+O\left[(R\omega_0)^2\right],\nn\\
\Gamma_{-}&=&O\left[(R\omega_0)^{2\lceil{\omega_0}/{\Omega}\rceil}\right],\hspace{0.5cm}
\eea
where $O\left[x^n\right]$ denotes that infinitesimals of the nth and higher orders of $x$ are omitted, and $\lceil x\rceil$ gives the smallest integer greater than or equal to $x$.
From Eq. \eqref{approximate-z}, it is obvious that \emph{1)} The excitation rate of the atom is always a higher-order small quantity compared with the emission rate and \emph{2)} When the orbital radius $R$ tends to zero, the emission and excitation rates of the atom tend to those of inertial atoms in vacuum. This aligns with one's intuition since the centripetal acceleration  $a=\gamma^2R\Omega^2$ vanishes when $R\to0$.

\emph{(II)\;Atomic polarization perpendicular to the axis of rotation}:
We now consider the transition rate contributed by the atomic polarization perpendicular to the axis of rotation (i.e., the radial and tangential direction of the orbit).
In this case, the approximated expressions of the emission and excitation rates in the limit of a small  orbital radius $R$ ($v=R\Omega\ll1$ and $R\omega_0\ll1$) are obtained from Eq. \eqref{Gammapm} as follows
\bea\label{approximate-r}
\Gamma_{+}&=&\left(d_{r}^2+d_{\theta}^2\right)\frac{\omega_0^3 +3\Omega^2\omega_0}{3\pi }+O[(R\omega_0)^2],\nn\\
\Gamma_{-}&=&O\left[(R\omega_0)^{2\lceil{\omega_0}/{\Omega}\rceil-2}\right],\hspace{0.5cm}
\eea
for $\Omega<\omega_0$, and
\bea\label{approximate-r2}
\Gamma_{\pm}&=&\left(d_{r}^2+d_{\theta}^2\right)\frac{(\Omega\pm\omega_0)^3}{6 \pi }+O\left[(R\omega_0)^2\right],
\eea
for $\Omega\geq\omega_0$. From Eqs. \eqref{approximate-r} and \eqref{approximate-r2}, we can draw the following  conclusions.

\emph{1)} When the rotational angular velocity does not reach the transition frequency of the atom, i.e, $\Omega<\omega_0$, the excitation rate is a higher-order small quantity compared with the emission rate.  As the orbital radius $R$ tends to zero, the excitation rate  approaches that of an inertial atom in vacuum, meaning it will vanish. In contrast, the emission rate does not vanish because the rotation  introduces an extra term $\left(d_{r}^2+d_{\theta}^2\right)\frac{\Omega^2\omega_0}{\pi }$ to the emission rate.

\emph{2)} When the rotational angular velocity exceeds the transition frequency of the atom, i.e, $\Omega>\omega_0$, the excitation rate will be of the same order of magnitude as the emission rate, even in the limit of a small orbital radius $R$ such that the centripetal acceleration $a=\gamma^2R\Omega^2$ is vanishing. 

\emph{3)}  Comparing Eqs. \eqref{approximate-r} and \eqref{approximate-r2} with Eq. \eqref{approximate-z}, one can find that, when the orbital radius is small, the excitation rate contributed by the atomic polarization along the direction of the axis of rotation is always an infinitesimal of higher order than that contributed by the atomic polarization perpendicular to the axis of rotation, regardless of whether the rotational angular velocity exceeds the transition frequency of the atom. 

\emph{The effective temperature}.---Usually, atoms  are isotropically polarizable. 
Since the atomic polarizability is related to the dipole transition matrix elements as 
$\alpha_{i}=2|d_i|^2/\omega_0$ for a two-level atom \cite{Friedrich1990}, and we have assumed that $d_i$ are real, then  isotropic polarizability implies that 
$d_r^2=d_{\theta}^2=d_z^2\equiv d^2$. Then, according to Eqs. \eqref{approximate-z}, \eqref{approximate-r} and \eqref{approximate-r2}, one obtains
\bea\label{sum-transl}
\Gamma_{+}&=&d^2\frac{\omega_0^3+2\Omega^2\omega_0}{\pi }+O\left[(R\omega_0)^2\right], \nn\\
\Gamma_{-}&=&O\left[(R\omega_0)^{2\lceil{\omega_0}/{\Omega}\rceil-2}\right],
\eea
for $\Omega<\omega_0$, and
\bea\label{sum-trans}
\Gamma_{+}&=&d^2\frac{(\Omega+\omega_0)^3+\omega_0^3}{3 \pi }+O\left[(R\omega_0)^2\right], \nn\\
\Gamma_{-}&=&d^2\frac{(\Omega-\omega_0)^3}{3 \pi }+O\left[(R\omega_0)^2\right],
\eea
for $\Omega\geq\omega_0$. 

As the spontaneous transition persists for a sufficiently long time, the ratio of the probabilities for the atom in the ground and excited states reaches a steady value, and the steady state of the atom $\rho(t)$ becomes a thermal state
\begin{equation}\label{rho_inf}
    \lim_{t\to\infty}\rho(t)=\frac{e^{-H_A/T_{\mathrm{eff}}}}{\mathrm{Tr}[e^{-H_A/T_{\mathrm{eff}}}]},
\end{equation}
where $H_A$ is the Hamiltonian of the two-level atom, and 
\begin{equation}\label{tem}
    T_{\mathrm{eff}}=\frac{\omega_{0}}{\ln(\Gamma_{+}/\Gamma_{-})},
\end{equation}
is the effective temperature. For an ensemble of rotating two-level atoms, the effective temperature characterizes the ratio of the population of atoms in the ground and excited states in the steady state. See, e.g., Refs. \cite{Bell83,Unruh98}. 
According to 
the definition of the effective temperature \eqref{tem} and Eqs.~\eqref{sum-transl}-\eqref{sum-trans}, when the rotational angular velocity is less than the transition frequency of the atom, i.e, $\Omega<\omega_0$, the effective temperature is
\beq
T_{\text{eff}}\sim\omega_0/\ln\left[(R\omega_0)^{2-2\lceil{\omega_0}/{\Omega}\rceil}\right],
\eeq
which depends on the orbital radius $R$, and vanishes as $R$ and thus the centripetal acceleration tend to zero. However, when the rotational angular velocity exceeds the transition frequency of the atom, i.e, $\Omega>\omega_0$, 
the effective temperature  becomes
\bea \label{effective-temperature}
T_{\rm eff}=\frac{\omega_0}{\ln\left\{{[(\Omega+\omega_0)^3+\omega_0^3]}/{(\Omega-\omega_0)^3}\right\}},
\eea
which is independent of the orbital radius and the centripetal acceleration. 
In principle, the orbital radius and thus the centripetal acceleration can be arbitrarily  small while keeping the rotational angular velocity larger than the transition frequency of the atom. 
Therefore, surprisingly, a large effective temperature in the centripetal acceleration case can be obtained at an almost vanishing centripetal acceleration. Furthermore, when the rotational angular velocity is much larger than the transition frequency, i.e., when $\Omega\gg\omega_0$, 
the effective temperature can further be approximated as
\bea
T_{\rm eff}\approx \frac{\Omega}{6}.
\eea
This shows that the effective temperature is approximately proportional to the angular velocity when the rotational angular velocity is much larger than the energy level spacing, which is remarkably  different from  the linear acceleration case, where the temperature is proportional to the acceleration.

\emph{Discussion}.---A few comments are now in order.

First, the reason that the transition rate for the rotating atom does not reduce to that in the inertial case in the limit of a vanishingly small orbital radius and thus a vanishingly small centripetal acceleration can be traced back to the interaction Hamiltonian~\eqref{hi2}. In  the limit  of $R\to0$ and thus $v\to0$ and $\gamma\to1$, but the angular velocity $\Omega$ is finite, the interaction Hamiltonian reduces to $H_I=-\sum_{i=r,\theta,z}D_{i}{E}_{i}(t,{\bf x})$. Although this seems to be the usual dipole interaction Hamiltonian, one should bear in mind that ${D}_{i}$ are components of the dipole operator in the proper frame of the rotating atom, which are related to those in the laboratory frame by a rotation transformation. For atoms polarizable along the direction of the axis of rotation, the dipole is invariant under the rotation transformation.  However,  this is not the case for atoms polarizable perpendicular to the direction of the axis of rotation.  
This explains why the transition rate for a rotating atom with a finite angular velocity does not reduce to that in the inertial case in the limit of a vanishingly small centripetal acceleration. Note that this phenomenon does not exist in the toy model in previous studies  where  the atom is assumed to be in the monopole interaction with a scalar field  \cite{Kim1987,Letaw1981,Korsbakken2004,Good2020,Takagi1984,Takagi1986,Biermann2020,Bunney23,Letaw1980,Levin1993,Davies1996,Lorenci2000,Gim2018,Bunney23-2}. 

Second, why does the excitation rate contributed by the atomic polarization perpendicular to the axis of rotation change dramatically when the rotational angular velocity $\Omega$ crosses the energy level spacing $\omega_0$ in the limit of a small orbital radius $R$? 
To answer this question, let us note that when the orbital radius $R$ is small such that $v=R\Omega\ll1$,
the field correlation functions \eqref{Wightman-correlation-function-rr-m}-\eqref{Wightman-correlation-function-zz-m} can be approximated as,
\bea
G_{rr}(\Delta t)&=&G_0(\Delta t)\frac{e^{i\Omega\Delta t}+e^{-i\Omega\Delta t}}{2}+O[(R\Omega)^2],\hspace{0.5cm}\label{Grr0}\\
G_{\theta\theta}(\Delta t)&=&G_0(\Delta t)\frac{e^{i\Omega\Delta t}+e^{-i\Omega\Delta t}}{2}+O[(R\Omega)^2],\hspace{0.5cm}\label{Gtt0}\\
G_{zz}(\Delta t)&=&G_0(\Delta t)+O[(R\Omega)^2],\label{Gtt0}
\eea
where $G_0(\Delta t)=\frac{1}{\pi^2(\Delta t-i \epsilon)^4}$ is the field correlation function obtained by taking the trajectory of an inertial atom into the general expression \eqref{field correlation function}, with $\epsilon$ being a positive infinitesimal. Now, 
one can easily obtain, in the leading order,  from Eqs. \eqref{Grr0}-\eqref{Gtt0} that, 
for the case of the atomic polarization along the radial or tangential direction of the orbit,
\bea\label{Gamma-frr}
\Gamma_{\pm}(\omega_0)=
\frac{1}{2}\left[\Gamma_{\pm}^{0}(\omega_0+\Omega)+\Gamma_{\pm}^{0}(\omega_0-\Omega)\right].
\eea
Here, $\Gamma_{+}^{0}(\omega_0)={d^2\omega_0^3}/{(3\pi)}$ and $\Gamma_{-}^{0}(\omega_0)=0$ are the spontaneous emission rate and excitation rate for inertial atoms respectively. 
Moreover, from Eq. \eqref{eer}, we have that
\bea
\Gamma_{\pm}(-\omega)=\Gamma_{\mp}(\omega). 
\eea
Therefore, for the case of the atomic polarization along the direction perpendicular to the axis of rotation, when the orbital radius is sufficiently small and the rotational angular velocity $\Omega$ is smaller than the energy level spacing $\omega_0$, i.e., $\Omega<\omega_0$ , the excitation rate is approximately half of the sum of the excitation rates of two inertial atoms in vacuum with the energy level spacing being $\omega_0+\Omega$ and $\omega_0-\Omega$ respectively, which is vanishing. However, when $\Omega>\omega_0$, i.e., $\omega_0-\Omega<0$, the excitation rate is approximately half of the sum of the excitation rate of an inertial atom with an energy level spacing $\omega_0+\Omega$ (which is vanishing), and the {\it emission} rate of an inertial atom with an energy level spacing $\Omega-\omega_0$ (which is finite). 
This explains why the excitation rate undergoes a sudden change when the rotational angular velocity $\Omega$ crosses the energy level spacing $\omega_0$. 

Third, the rotating atoms are treated as pointlike particles in this Letter. However, actually, the size of an atom must be nonvanishing. In order to avoid the consideration of the finite-size effect,   the orbital radius $R$ should be much larger than the radius of the atom characterized by the Bohr radius $R_B\sim10^{-10}$ m, although it has  been assumed to be small such that $R\omega_0\ll1$ and $R\Omega\ll1$. 
Note that the typical energy level spacing of an atom is of the order of $\omega_0\sim1$ eV, which corresponds to a transition wavelength $\omega_0^{-1}\sim10^{-6}$ m. Therefore, if we assume that the rotational angular velocity $\Omega$ is of the same order of $\omega_0$, there exists  plenty of parameter space for the orbital radius $R$ such that it is much larger than the radius of the atom and at the same time much smaller than the length characterized by the rotational angular velocity $\Omega$ and the energy level spacing  $\omega_0$.

Fourth, let us note that the transition rates shown in Eq.~\eqref{eer} are derived under the Markov approximation. By adopting the method presented in Ref.~\cite{Anastopoulos}, it can be shown that the Markov approximation is justified regardless of whether the rotational angular velocity $\Omega$ exceeds the transition frequency  $\omega_0$. See Sec.~\textcolor{blue}{IV} of the Supplemental Material for details.

Finally, let us discuss the possibility of verifying this effect in experiment. 
Recently, there have been efforts to achieve hyperfast rotation in an optically levitated nanoparticle system \cite{Reimann2018,Li2018,Zhang2021}, and the highest rotation angular frequency obtained so far is $\Omega\approx6$ GHz for nanoparticles with an average radius $R=95$ nm \cite{Zhang2021}.  Then, a natural idea is that we can manage to attach the atom to a hyperfast rotating nanoparticle, as proposed in Ref. \cite{Lochan20}. Based on the data in Ref. \cite{Zhang2021}, we assume that  $\omega_0\sim1$ GHz, $\Omega=6\omega_0$, $R\sim10^{-7}$ m, corresponding to a centripetal acceleration $a\sim 10^{11}\;\text{m}/\text{s}^2$.  This is much smaller than the transition frequency of the atom since $a/\omega_0\sim10^{-6}\ll1$. 
For a uniformly accelerated atom at such an acceleration, the excitation rate is 
completely negligible compared with the emission rate, since 
$\Gamma_-^{\text{linear}}\sim10^{-272876}\Gamma_+^{\text{linear}}$.
For a rotating atom, in contrast, the excitation rate is of the same order of the emission rate, 
and at the same time it is extraordinarily larger than that in the linear acceleration case, as now 
$\Gamma_-^{\text{circular}}\sim\Gamma_+^{\text{circular}}\sim10^{272878}\Gamma_-^{\text{linear}}$. This shows that, compared with the linear Unruh effect, the circular  Unruh effect can be significant even at a centripetal acceleration extremely small compared with the transition frequency of the atom. In fact, the excitation rate will be even larger if a larger rotational angular velocity can be reached.
For an optically levitated nanoparticle system, the rotational angular velocity increases as the size of the particle decreases \cite{Li2018}.  Moreover, from Eq.~\eqref{sum-trans}, we can see that the leading term of the transition rate is independent of the orbital radius $R$ when the angular velocity exceeds the energy level spacing. This indicates that, one can achieve a higher rotational angular velocity by reducing the size of the nanoparticle to which the atom is attached, thereby obtaining a more significant circular Unruh effect. Therefore, the phenomenon discovered in this Letter may potentially be observed with the help of an optically levitated nanoparticle system.

\emph{Summary}.---In this Letter, we have investigated the transition rate of an atom rotating in a circular orbit, which is coupled with fluctuating electromagnetic fields in vacuum. We have shown that when the rotational angular velocity exceeds the transition frequency of the atom, the spontaneous excitation rate, even in the limit of a vanishingly small orbital radius and thus a vanishingly small centripetal acceleration, is not only nonvanishing but also unexpectedly of the same order of magnitude as that of the emission rate.  Moreover,  the effective temperature of the circular Unruh effect, defined by the ratio of  the excitation and emission rates, is approximately proportional to the rotational angular velocity when the orbital radius is vanishingly small and the rotational angular velocity is much larger than the energy level spacing.  This is remarkably different from the linear acceleration case, where the temperature is proportional to the acceleration. Our result, therefore, suggests that a significant circular Unruh effect can be observed at a vanishingly small centripetal acceleration. This phenomenon may potentially be observed with the help of an optically levitated nanoparticle system.

\emph{Acknowledgments}.---We would like to thank Wenting Zhou for helpful discussions. We also extend our gratitude to the anonymous referees for their insightful comments and helpful suggestions. 
This work was supported in part by the NSFC under Grants No. 12075084 and No. 12375047, the innovative research group of Hunan Province under Grant No. 2024JJ1006, the Hunan Provincial Natural Science Foundation of China under Grant No. 2023JJ40515, and the Scientific Research Program of Education Department of Hunan Province of China under Grant No. 22B0762.

\end{document}